# Detection of COVID-19 in Chest X-Ray Images Using Transfer Learning

Zanobya. N. Khan

Department of Computer Science, SUIT, Peshawar, Pakistan
zanoby.nisar@gmail.com

**Abstract**

*COVID-19 – a highly contagious disease infected millions of people worldwide. With limited testing components, screening tools such as chest radiography can assist the clinicians in the diagnosis and assessing the progress of disease. The performance of deep learning-based systems for diagnosis of COVID-19 disease in radiograph images has been encouraging. This paper investigates the concept of transfer learning using two of the most well-known VGGNet architectures, namely VGG-16 and VGG-19. The classifier block and hyperparameters are fine-tuned to adopt the models for automatic detection of Covid-19 in chest x-ray images. We generated two different datasets to evaluate the performance of the proposed system for the identification of positive Covid-19 instances in a multiclass and binary classification problems. The experimental outcome demonstrates the usefulness of transfer learning for small-sized datasets particularly in the field of medical imaging, not only to prevent over-fitting and convergence problems but also to attain optimal classification performance as well.*

## 1. Introduction

The emergence of novel coronavirus at the end of December 2019 spread rapidly to other countries of the world and soon become a global pandemic. This infectious disease caused by severe acute respiratory syndrome coronavirus 2 (SARS-CoV-2) and named as COVID-19 by the International Committee on Taxonomy of Viruses, is now a major health concern worldwide. The SARS virus in 2003 infected 8093 people in 26 countries [1] whereas MERS infected 2428 individuals in 2012 [2]. The COVID-19, on the other hand, has affected more than four million people and caused 328,120 deaths as of May 21 [3]. These statistics show that SARS-CoV2 has a higher transmission rate than the previous two coronaviruses. COVID-19 is transmitted usually through respiratory droplets of an infected person or close contact with the one who is infected. It is a highly contagious virus [4] and people remain in incubation for two weeks or more. Lungs is the main organ attacked by the virus. This is the cause that some severely infected patients gradually develop respiratory [5][6].

Presently, reverse transcription-polymerase chain reaction (RT-PCR) test used as a primary tool for early diagnosis of COVID-19. But in some areas, testing resources are limited and the infected cases are rising [7]. As such, chest radiography can serve as a potential screening tool for diagnosis [8] and also to assist clinicians in assessing the progress of the disease among patients.

Over the years, deep learning techniques [9] have been used in various fields of science and industry. Convolutional neural networks deemed as the most successful type of deep learning algorithms achieved state-of-the-art performance in object detection and recognition, speech recognition, satellite imaging, natural language processing and biomedical imaging [9]. In 2012, the AlexNet model described by Krizheyzky et al. [10] won the ImageNet Large-Scale Visual Recognition (ILSVR) Challenge. This is considered an essential breakthrough in deep learning leading to its immediate adoption by the computer vision community [9]

In the field of medical imaging, deep convolutional neural networks are difficult to train. First, due to non-availability of sufficient data as medical images have to be manually examined and labeled by radiologists which can be costly and a time-consuming task. Moreover, training deep CNNs on small-sized datasets can lead to overfitting and convergence problems [12].

The hierarchical architecture of deep CNNs usually allows them to extract features at multiple levels. At the initial layers, features such as edges, color, or orientation are learned, but as we go deeper in the network more specific features are extracted. These general-specific feature transformations resulted in the development of transfer learning [13][14]. Thus, a more favorable alternative for a limited number of training samples is to use transfer learning that would achieve significant performance improvement over the CNNs trained from scratch.

In this work, we propose pretrained VGG-16 and VGG-19 for the detection of Covid-19 samples in chest x-ray images. The pre-trained networks are fine-tuned to solve multiclass and binary classification problems. Therefore, we created two datasets to conduct experiments and analyze the performance of pretrained models for both the \tasks. The contributions of this work include:

1. Two different datasets employed to perform multiclass and binary classification task
2. For improved results, transfer learning is performed by fine-tuning the pre-trained CNNs

## 2. Related Works

Narin et al. [15] used pre-trained InceptionV3, ResNet50 and Inception-ResNetV2 models for Covid19 detection. They used a dataset of 100 images containing 50 normal and 50 Covid19 infected chest X-ray images and obtained an accuracy of 98.0% with ResNet50 architecture. A ResNet based model, trained using progressive resizing and data augmentation achieved an overall accuracy of 0.962 on COVIDx dataset [16]. The ResNet model propose by Zhang et al. [17] performed two tasks: a binary classification of the chest X-ray images as COVID-19 and non-COVID-19 and detection of anomalies. The model was able to detect 96% COVID-19 and 70.65% non-COVID-19 samples on a dataset with 100 COVID-19 images and 1431 pneumonia images. The DarkCovidNet designed for multi-class and binary classification tasks achieved a classification accuracy of 87.02% and 98.08% respectively. The authors evaluated the performance of their proposed system using K-fold cross-validation with 125 COVID-19 images, 500 normal, and pneumonia chest X-ray images [18]. The SqueezeNet model with Bayes optimization and offline augmentation by Ucar and Korkmaz [19] achieved a test accuracy of 0.983 on the COVIDx dataset. The COVIDx dataset consists of 76 COVID-19 samples, 1583 normal, and 4290 pneumonia samples. The COVID-Net system designed by Wang and Wong [20] intended to classify images into four classes, namely COVID19, normal, viral pneumonia and bacterial pneumonia. Setha and Behera [21] extracted deep features from several deep learning algorithms and used Support Vector Machine (SVM) as a classifier. However, an accuracy of 95% was achieved only with ResNet50 among other models.

## 3. Materials and Methods

### 3.1 Datasets

In this research, the proposed scheme will be evaluated for a multiclass and binary classification scenario. For this purpose, we build two datasets referred to as Dataset-1 and Dataset-2. Both the dataset contains 278 Covid-19 x-ray images with 223 samples curated from the open-source GitHub repository [22] and the remaining 55 samples from [23]. Pneumonia (including bacterial and viral pneumonia) and normal images were obtained from the Kaggle Chest X-Ray Images repository [24]. For a binary classification task, non-COVID-19 images were collected from [25]. In addition to normal and pneumonia images, includes images of other lungs diseases as well. The distribution of images in the two datasets is given in Table I.

Table 1: Class distribution of images in the datasets

| Dataset | | No. of Images |
|---|---|---|
| Dataet-1 | COVID-19 | 278 |
| | Non Covid-19 | 978 |
| Dataset-2 | COVID-19 | 278 |
| | Pneumonia | 776 |
| | Normal | 1120 |

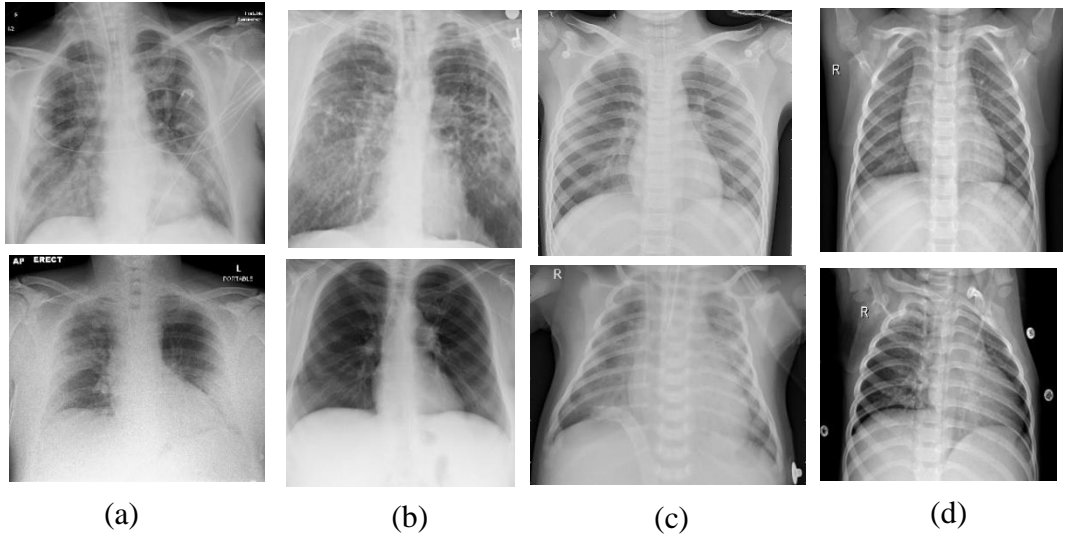

| (a) | (b) | (c) | (d) |

Figure 1: Randomly selected X-ray images (a) COVID-19 (b) non-COVID-19 (c) pneumonia (d) normal

## 3.2 Evaluation Metrics

In order to assess the performance of the proposed method four different statistical measures precision, recall, F-measure and accuracy coefficient are adopted. These metrics are mathematically expressed as:

$$Precision = \frac{TP}{(TP + FP)} \quad (1)$$

$$Recall = \frac{TP}{(TP + FN)} \quad (2)$$

$$F - measure = 2 \times \frac{precision \times recall}{(Precision + Recall)} \quad (3)$$

$$Accuracy = \frac{TP + TN}{(TP + FP + FN + TN)} \quad (4)$$

## 3.3 Transfer Learning

In transfer learning, a commonly adopted approach is training a deep CNN on large-scale labeled data, such as ImageNet, and then transfer the pre-trained network to a small dataset referred to as target data. For medical image analysis, two different transfer learning strategies are generally adopted. In the first approach, the pretrained CNNs are used as feature extractors and these features are then passed to linear classifiers such as SVM for classification. The second approach is fine-tuning, where the fully connected layers of pre-trained CNNs are replaced with one or more new layers and according to the target domain and re-trained with the newly added layers [26].

## 3.4 VGGNet Architecture

In this work, we apply fine-tuning to two-well known deep learning models namely VGG-16 and VGG-19 trained on large-scale ImageNet dataset [11]. It was proposed by K. Simonyan and A. Zisserman at the University of Oxford [27] and was placed 1st in localization and 2nd in classification at ImageNet Large Scale Visual Recognition

Challenge (ILSVRC) 2014. The VGGNet is trained on 1.2 million images with 1000 different classes. VGG-16 contains 13 convolutional layers, three fully-connected layers, and five max-pooling layers. Each convolutional block consists of 2D convolutional layers with kernels of size 3x3 and stride one followed by max-pooling layers. Spatial pooling is carried out by max-pooling layers with kernels of size 2x2 and a stride of two. Each convolutional layer is followed by an activation function, namely rectified linear unit (ReLu). The classifier block is composed of three fully connected layers with 4096 neurons in the first and second layer, and the last layer with softmax activation function contain 1000 neurons. VGG-19, on the other hand, contains 16 convolutional layers, three fully-connected layers, and five max-pooling layers. The use of 3x3 filters in all the convolutional layers with strides of size one and it's the promising performance with limited training data in the field of medical imaging motivated us to use VGGNet [27].

### 3.5 Proposed CNN Architecture for COVID-19 Detection

For COVID-19 detection in CXR images, we modified the classifier block of VGG-16 and VGG-19 according to our problem domain. The input images are rescaled to a size of 192x192. The last two fully-connected layers composed of 512 neurons instead of 4096 neurons as proposed in the original architecture[27], trained on the large-scale ImageNet dataset. A dropout layer is used after every fully-connected layer to avoid overfitting. The last fully-connected layer is removed and replaced with the layer suitable for CXR datasets.

For multi-class classification, the neurons in the last fully-connected layer is set to three. On the other hand, the sigmoid activation function with two neurons is used in the last fully-connected layer for a binary classification task. The size of the input images and configuration of the fully-connected layers are same in the two networks. Figure 2 depicts the proposed framework of the multiclass scenario.

We also fine-tuned different hyperparameters to achieve the desired outcome. These include learning rate, batch size, and epochs. Both the models compiled using an Adam optimizer with a learning rate of 1e-4. Other parameters such as batch size and epochs are dataset specific and listed in Table II. The fine-tuning of the last three fully-connected layers and hyperparameters in both the networks resulted in the effective detection of COVID-19 samples in CXR images.

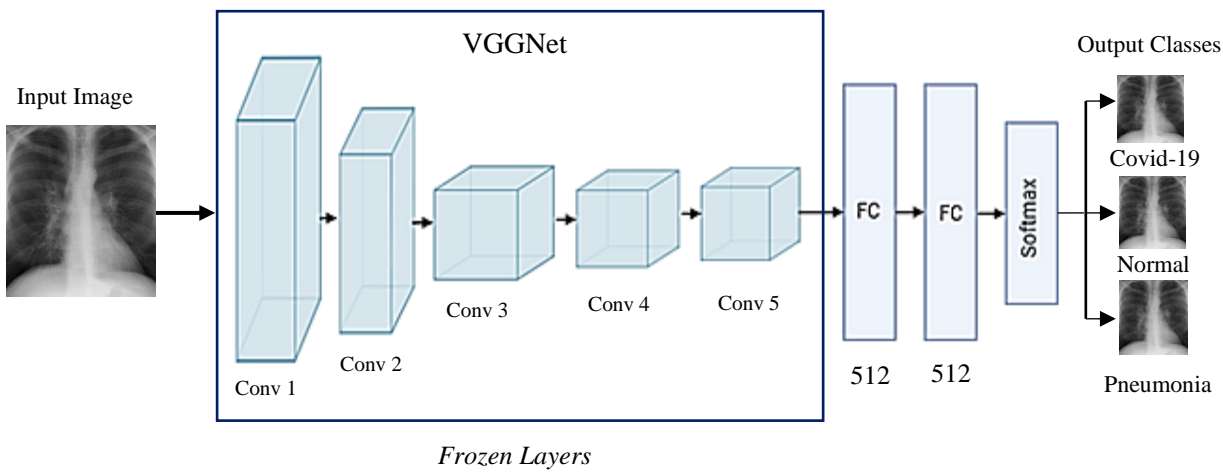

Figure 2: Proposed framework for multiclass classification of x-ray images

Table 2: Hyperparameters setting for the classification tasks

| | |
|---|---|
| Dataset-1 | Batch Size = 24 |
| | Number of Epochs = 12 |
| Dataset-2 | Batch Size = 32 |
| | Number of Epochs = 16 |

## 3.7 Experimental Setup

In this study, both the datasets were partitioned into a training dataset, a validation dataset, and a test dataset. The training dataset contains 80% of total images and the remaining 20% used to evaluate the classification performance. The training data is further split using a ratio of 70:10 into training and validation datasets. To avoid over-fitting, data augmentation operations of rotation and flipping are applied to both the training and validation sets.

## 4.1 Experimental Results

In this research, several experiments are conducted to assess the effectiveness of transfer learning and fine-tuning for Covid-19 detection in CXR images. At first, we provide a comparative analysis of the binary classification problem by performing experiments on Dataset-1. And finally, the performance of the proposed scheme for a multiclass scenario on Dataset-2 is evaluated.

Table 3 presents the experimental outcome of pre-trained models fine-tuned for a binary classification task. The result indicates an unexpected low performance of VGG-19 (0.924 accuracy) as compared to VGG-16. Furthermore, the misclassification rate for non-COVID-19 samples is also high as depicted in Figure 3. VGG-16, on the other hand, achieved precision, recall, F-measure, and accuracy of 0.915, 0.935, 0.930, and 0.948 on the test dataset.

Further, the confusion matrix in Figure 3 reveals that VGG-16 successfully classified 51 COVID-19 samples as against the 48 by VGG-19 from total 56 samples in the test dataset. Perhaps this can be due to the shallower depth of VGG-16 model in contrast to the VGG-19 architecture. Although VGG-19 should generally yield better classification performance, however the relatively small size of our training data and dense architecture, probably overfit to the limited data, whereas the shallower model generalized better in this scenario.

In the next phase, experiments are carried out to assess the performance of the system for a multiclass classification problem. It is evident from the results in Table 4 that there is a slight difference in the performance measure of VGG-16 and VGG-19 in a multiclass scenario. VGG-16 obtained a classification accuracy of 0.928 as opposed to 0.917 by VGG-19. Precision, recall, and F-measure values also don't vary greatly.

To further analyze the performance of the system corresponding to the identification of Covid-19, the confusion matrix is presented in Figure 4. As shown in Figure 4, VGG-19 correctly classified most of the Covid-19 samples with only one misclassified instance. Furthermore, VGG-19 produced improved results in distinguishing between normal and pneumonia cases than the VGG-16 model. An improvement in the prediction accuracy of VGG-19, specifically for Covid-19, can be attributed to the size of training data. Besides, deep learning models greatly enhance the classification performance when trained with large amounts of data. Comparatively, VGG-16 misclassify two positive Covid-19 samples and proved to be more successful in discriminating pneumonia cases from normal, and this contributed to an overall better classification performance.

We can conclude from the results that transfer learning and fine-tuning the pre-trained deep CNNs can be more effective than end-to-end training specifically for small-sized datasets as in our scenario and biomedical imaging field in general.

Table 3: Performance of binary classification on the test set of Dataset-1

| Model  | Precision | Recall | F-Score | Accuracy |
|--------|-----------|--------|---------|----------|
| VGG-16 | 0.915     | 0.935  | 0.93    | 0.948    |
| VGG-19 | 0.88      | 0.90   | 0.89    | 0.924    |

Table 4: Performance of multiclass classification on the test set of Dataset-2

| Model | Precision | Recall | F-Score | Accuracy |
|---|---|---|---|---|
| VGG-16 | 0.939 | 0.934 | 0.936 | 0.928 |
| VGG-19 | 0.931 | 0.924 | 0.926 | 0.917 |

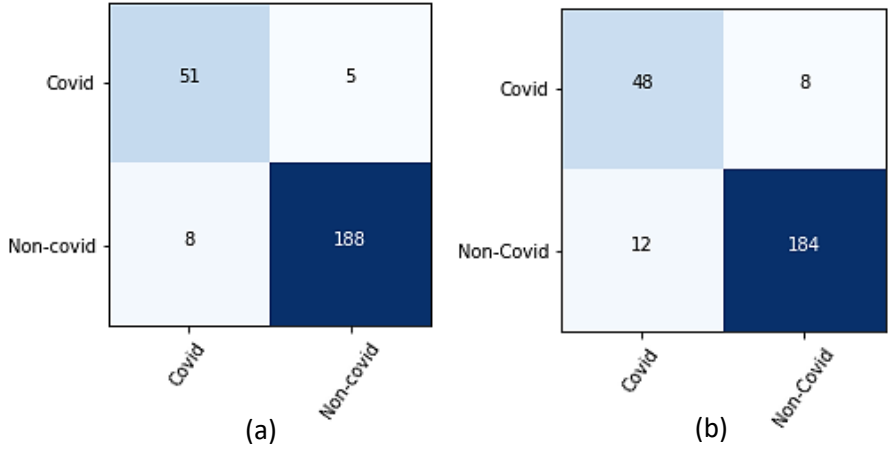

Figure 3: Confusion matrix for binary classification (a) pretrained VGG-16 (b) pretrained VGG-19

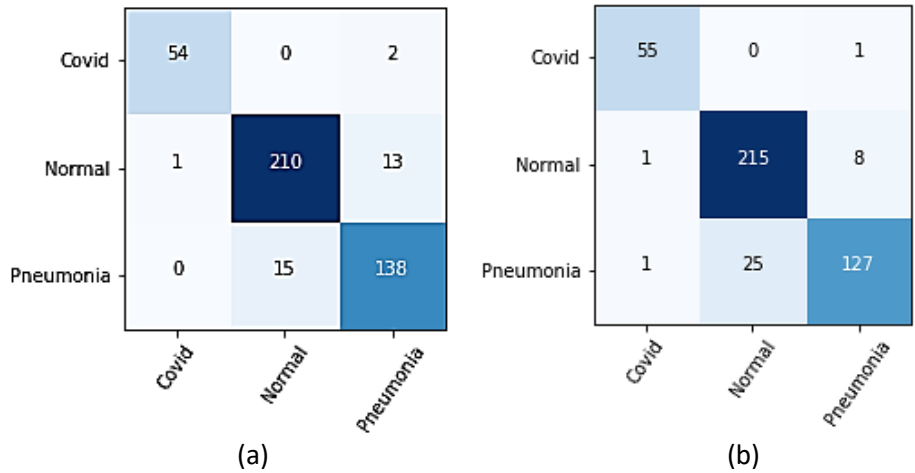

Figure 4: Confusion matrix for multiclass classification (a) pretrained VGG-16 (b) pretrained VGG-19

### 4.2 Discussion

As no benchmark dataset is yet available to analyze COVID-19 cases in chest x-ray images, authors collected images from different sources to build the datasets and conduct experiments. Table 5. presents comparison of the proposed framework with state-of-the-arts. A deep learning-based COVID-Net model designed by Wang and Wong [20] acquired an accuracy of 92.4% with 53 COVID-19 samples out of the 16,756 total images. Ozturk et al. [18] tested the proposed DarkCovidNet by using 125 COVID-19, 500 pneumonia, and 500 normal images and achieved 98.08% and 87.02% for binary and ternary class scenario. Ioannis et al. [28] used deep transfer learning for the detection of COVID-19 in two datasets. The first dataset comprises 224 COVID-19, 504 normal, and 700 bacterial pneumonia images and achieved an overall accuracy of 98.75% and 93.48% for 2-class and 3-class respectively with the VGG-19 model. The second dataset contains 224 COVID-19, 504 normal and, 714 pneumonia includes both viral and bacterial images. And for 2-class and 3-class scenarios achieved an accuracy of 96.78% and 94.72% using pretrained MobileNet-V2. In the proposed work, however, we used different datasets for 2-class and 3-class classification problem and achieved an accuracy of 94.8% and 92.8% by fine-tuning the pre-trained VGG-16 model.

Table 5: Comparison of proposed work with state-of-the-art

| Authors | Number of Images | Method | Accuracy |
| --- | --- | --- | --- |
| Wang & Wong [20] | COVID-19: 53<br>Pneumonia: 5826<br>Normal: 8066 | COVID-Net | 92.4% |
| Ozturk et al. [18] | COVID-19: 125<br>Pneumonia: 500<br>No-Findings: 500 | DarkCovidNet | 87.02% |
| Ioannis et al. [28] | COVID-19: 224<br>Pneumonia: 700<br>Normal: 504 | VGG-19 | 93.48% |
| | COVID-19: 224<br>Pneumonia: 714<br>Normal: 504 | MobileNet-V2 | 94.72% |
| Proposed | COVID-19: 278<br>Pneumonia: 776<br>Normal: 1120 | VGG-16 | 92.8% |
| | COVID-19: 278<br>Non COVID-19: 978 | VGG-16 | 94.8% |

**Conclusion**

This paper presents pre-trained VGG-16 and VGG-19, fine-tuned to successfully identify positive Covid-19 samples in chest x-ray images. We evaluated performance of the proposed framework in different classification scenarios. In a binary classification task, VGG-16 showed promising results in the detection of Covid-19 samples whereas in multiclass problem, VGG-19 correctly classified majority of the Covid-19 cases. However, in both the scenarios, VGG-16 achieved an overall better classification performance. The results obtained in both the scenarios, suggests that fine-tuning the pre-trained deep learning architectures can generate better performance with limited training data rather than training the entire network from scratch.